\documentstyle[11pt]{article}
\newcommand{\blankline}{\vskip .3cm}
\newcommand{\f}{\begin{equation}}
\newcommand{\ff}{\end{equation}}

\begin{document}
\rightline{cgpg-94/3-5}
\blankline
\centerline{\LARGE The fate of  black hole singularities}
\blankline
\centerline{\LARGE and}
\blankline
\centerline{\LARGE the parameters of the standard models}
\centerline{\LARGE of particle physics and cosmology}
\blankline
\rm
\centerline{Lee Smolin${}^*$}
\blankline
 \centerline{\it  Center for Gravitational Physics and Geometry}
\centerline{\it Department of Physics}
 \centerline {\it The Pennsylvania State University}
\centerline{\it University Park, PA, USA 16802}
 \vfill
\centerline{March 28, 1994}
\vfill
\centerline{ABSTRACT}
\blankline
\noindent
The implications of a cosmological scenario which explains the values
of the parameters of the standard models of  elementary particle
physics and cosmology are discussed.  In this scenario these
parameters are set by a process analogous to natural selection which
follows naturally from the assumption that the singularities in black
holes are removed by quantum effects leading to the creation of new
expanding regions of the universe.  The suggestion of J. A. Wheeler
that the parameters change randomly at such events, leads naturally to
the conjecture that the parameters have been selected for values that
extremize the production of black holes. This leads directly to a
prediction, which is that small changes in any of the parameters
should lead to a decrease in the number of black holes produced by
the universe.  Thus, in this case a hypothesis about particle physics
and quantum gravity may be refuted or verified by a combination of
astrophysical observation and theory.

This paper reports on attempts to refute this conjecture.  On
plausible astrophysical assumptions it is found that changes in many
of the parameters do lead to a decrease in the number of black holes
produced by spiral galaxies.  These include the masses of the proton,
neutron, electron and neutrino and the weak, strong and electromagnetic
coupling constants. Finally,  this scenario predicts a natural time scale
for cosmology equal to the time over which spiral galaxies maintain
appreciable rates of star formation, which is compatible with current
observations that $\Omega = .1-.2$.
\blankline
${}^*$ smolin@phys.psu.edu
\eject

\section{Introduction}

One of the great puzzles of astronomy and physics is what happens
inside of black holes, where general relativity breaks down because
of the presence of singularities\cite{bh-reviews}.  That
this is not just a problem of
mathematical physics is apparent if one reflects on the fact that
the rate of formation of black holes in the observable universe is
likely to be as high as one hundred per second\footnote{This
estimate is gotten by multiplying the $10^{12}$
galaxies believed to be
within our horizon by the rate of type II supernovas of about one
every forty years per spiral galaxy, and a conservative estimate that
between $1/10$ and $1/100$ of these become black
holes\cite{Shapiro-T}.}, this may
be taken to be the rate at which our ignorance about the universe is
increasing due to our not knowing what lies behind all of these event
horizons.   When one adds quantum
physics to the picture the puzzle becomes a crisis, as first realized
by
 Hawking in 1974, because of the problem of the loss of information
constituting the quantum state of the star whose collapse formed
the
black hole\cite{hawking-bhe}.

Another basic problem of physics is to understand why the masses
and coupling constants of the elementary particles take the values
they do.  This mystery, which has stubornly resisted solution
despite enormous progress in our understanding of the fundamental
interactions, is deepened when one tries to understand
why so many of the fundamental dimensionless constants that
describe the masses and interactions strengths are very large or
very
small numbers.  It is even further deepened when it is pointed out
that the
fact that our universe is as structured as it apparently is, from
the scales of galaxies to the existence of many stable nuclei, and
hence stars and chemistry, is based on a series of apparent
coincidences relating the values of the fundamental dimensionless
parameters of physics and cosmology.  For example, if one requires
that main sequence stars exist then (as will be outlined shortly) one
constrains the values of the
following quantities: the proton neutron mass difference, the
electron-nucleon mass ratio, $\alpha$, the strong interaction
coupling
constant, the neutrino mass\cite{CarrRees,BarrowTipler}.  Further
requiring that there are type II
supernova fixes a relation between the weak interaction and
gravitational constant given by eq. (7)
below\cite{BarrowTipler,CarrRees}
while requiring that there are convective stars fixes a relation
between the gravitational constant and the
fine structure constant\cite{Carter,BarrowTipler}
given by eq. (6).    With
these relations essentially every dimensionless constant associated
with the properties of stable matter has been fixed by the
requirement that stars with the life cycle of those in our universe
exist.

The purpose of this paper is to present evidence for a
cosmological conjecture that
relates these two puzzles\cite{evolve}.
The conjecture is simple to state, and
is a natural outgrowth of  ideas which have been contemplated
by particle physicists and relativists for many years.  As I will
describe, it leads to a definite and testable
prediction, which is that,

$*$ {\it  Almost every small change in the parameters of
the standard models of particle physics and cosmology
will either result in a universe
that has less black holes than our present universe,
or leaves that number unchanged.}

After I motivate it, the bulk of this paper will be devoted to
presenting evidence in favor of this prediction.

\section{Cosmological natural selection}

A natural solution to the problem of the fate of black hole
singularities, that has been discussed for
many years\footnote{I learned of it from Bryce DeWitt in 1980,
but I do not know who was the first to discuss it.},
 is that quantum effects cause a bounce when densities
become extreme (presumably of order of the Planck density) so that
the worldlines of the stars atom that have been converging begin to
diverge.   As there is nothing that can remove the horizon, before, at
least, the evaporation time of the black hole, which is at least
$10^{54}$ Hubble times for an astrophysical black hole and therefor,
plausibly, beyond the scope of this paper, whatever new region of
spacetime is traced by these diverging geodesics remains hidden
behind the original horizon.  Moreover, any observers in this new
region see themselves to be in a region of spacetime which is locally
indistinguishable from an expanding cosmological solution with an
apparent singularity in the past of every geodesic.  Thus, it would
make sense to call this process the creation of a new universe that
is
(at least on scales shorter than $10^{54}$ Hubble times) causally
disconnected from our universe\footnote{I will use here the
informal expression "universe" to mean a causally connected region
of spacetime, bounded by event horizons and excluding
any region where the
density of energy or curvatures approach Planck scales.  Roughly
speaking it corresponds to a region in which the laws of classical
general relativity may be relied on.}.

It may then be conjectured that each black hole of our universe leads
to such a creation of a new universe and that, correspondingly, the
big bang in our past is the result of the formation of a black hole in
another universe.

To have a theory of what determines the parameters of particle
physics and cosmology we need add only one equally natural
postulate to this picture.  It has been suggested a long time ago by
Wheeler\cite{JAW},
 and perhaps others, that the parameters of physics and
cosmology can change at such initiations of universes.  Let us
make the more specific assumption that {\it all the dimensionless
parameters of the standard models of particle physics and
cosmology change by small random increments at such
events.}\footnote{We may note that this is consistent with
our present understanding
of string theory and grand unified models of various kinds, as it
typically happens in these theories that the parameters of
the standard model that describes low energy
physics are determined by a particular solution of the more
fundamental theory.  What we need from such a theory to justify
the assumptions made here is that there is a large space of
solutions to the fundamental theory leading to different low energy
physics, and that generically different solutions differ by small
changes in the low energy parameters.  It then may be that the
fundamental theory will predict that when a region of the universe
approaches Planck densities there can be transitions between these
different solutions of the fundamental theory.  That this may be
possible is certainly consistent with what is presently known
about string theory.}

Then we have the following picture.  If we let $\cal P$ be the space
of dimensionless parameters, $p$, then  we can define an ensemble
of universes by beginning with an initial value $p^*$ and letting
the system evolve through $N$ generations.  Let us define a
function $B(p)$ on $\cal P$ that is the expected number of future
singularities generated during a lifetime of a universe with
parameters\footnote{We may note that even if the universe is open it is
very unlikely that the number of black holes produced during its
lifetime is infinite.  Thus, it is not necessary to make the assumption
made in \cite{evolve} that the universe is closed.  This was pointed
out by \cite{EllisRothman}} $p$.
We may observe
that, for most $p$, $B(p)$ is one, but there are small regions
of the parameter space where $B(p)$ is very large.  The present
values of the parameters must be in one such region because
there are apparently
at least $10^{18}$ black holes in our universe.

After $N$ generations
the ensemble then defines a probability distribution function
$\rho_N (p)$ on $\cal P$.  To give meaning to the postulate that
the random steps in the parameter space are small,  we
may require that the mean size of the random
steps in the parameter space is small compared to the width
of the peaks in $B(p)$.  It then follows from elementary
statistical configurations that,
for any starting point $p^*$ there
is an $N_0$ such that for all $N>N_0$, $\rho_N (p)$ is concentrated
around local maxima of $B(p)$.  This is because (from the
above restriction on step size)  it is overwhelmingly
probable that a universe picked at random from the ensemble is
the progeny of a universe that had itself many black holes.  But,
again, because the parameters change by {\it small} amounts at each
almost-singularity this means that it is overwhelmingly probable
that a universe picked at random from the ensemble itself has many
black holes.  Thus, we conclude that
{\it a typical universe in the ensemble (for $N>N_0$) has parameters
$p$ close to a local maximum of $B(p)$. }

Thus, the statement $*$ follows from the postulates we have
made concerning the fates of stars that collapse to black holes.

We may note that this theory is much stronger than any version of
the anthropic
principle\cite{Carter,CarrRees,BarrowTipler} because it conjectures
the existence of an
actual ensemble of universes that is generated by a
specific  process.  As a result, it necessarily
predicts that a certain
property must be satisfied by almost every universe in
the ensemble.  Furthermore, whether this property is true or false
of our
universe is determinable from physics and astrophysics at
observable scales.  Thus, this theory is highly vulnerable to
falsification.  This property  is not shared by any version of
the anthropic principle, first because there is no
principle that defines the ensemble in question and
second  because it requires only that there
exists in  whatever ensemble is conjectured only one universe
with a particular property, which is that there is intelligent
life\footnote{There is a recent proposal of
Crane according to which the anthropic principle would
become a consequence of the theory
discussed here if it happens often enough that
intellegent life
desires to, and is able to, construct black
holes\cite{anthrolouis}. This makes the anthropic principle a
particular hypothesis about cosmological natural selection in
the same way that one may discuss the selective advantage of
intellegence in biological natural selection.
Similarly, Crane's proposal is a scientific proposal, but
because life cannot evolve in a universe without galaxies
and stars, it
is one that cannot be discussed unless and untill the hypothesis
$*$ has been substantiated.  The above
comments then only refer to the anthropic principle prior to
such a discussion of
Crane's proposal.}.
A theory that asks that only one member of an (ill-defined, and
possibly infinite) ensemble exist with a
particular property can have no predictive
power, because it is possible that a member with
any set of logically possible properties
exist in such an ensemble\footnote{To my knowledge, the first
proposal that the quark masses and other parameters of the
standard
model might
be explained by a process analogous to natural selection
was made by Y. Nambu\cite{nambu}, although the
philosopher Charles Sanders Peirce made
similar speculations in the late
nineteenth century\cite{peirce}. More discussion of
the motivation behind the hypothesis of cosmological
natural selection, as well as more about its relationship
to the anthropic principle, may be found in ref \cite{book}.
The present paper is devoted only to discussion of the
testability of the conjecture.}

The  theory presented here
makes certain assumptions about physics at the Planck
scale which, presumably, may be tested directly at some time
in the future when we have a good understanding of that domain.
However, note that in order to test the prediction $*$, we need to
assume no more about Planck scale physics than was needed to
derive that statment.
Further, because there are many
dimensionless
parameters in the standard models of physics and cosmology,
and because so many of them are very small or very large, it is
easy to imagine that the statement $*$ could easily
be falsified without having to be very specific about the
width of the probability distributions around local maximum, its
dependence on $N$ or
any details of the form of $B(p)$.

Furthermore, because the argument
leads to a conclusion only about local maxima of $B(p)$ the
prediction $*$ refers to only
small changes in the parameters; it is irrelevant whether or not
there are parameters of $p$ very different from the present
values that lead to more black holes than are produced
by our present universe.

In the remainder of this paper I will discuss the evidence
for the statement $*$.

\section{Evidence for the prediction $*$}

As the standard models of physics and cosmology have about $20$
parameters, there are as many chances to falsify $*$.  At the
present time, the situation seems to be the following.  i)  $N(p)$ is
strongly sensitive to every cosmological parameter and to
every particle physics parameter that determines the
properties of stable matter.   ii)  No argument has so far been
found for
a small change in any parameter leading to an
increase in the number of black holes produced in the universe.
iii) Given reasonable and widely believed assumptions about
star formation processes in spiral galaxies there are clear
arguments
that at least seven distinct
small changes in the parameters that determine low energy
physics lead to a decrease
of
$N(p)$. These include changes in each of the four masses
of the stable particles: proton, neutron, electron and neutrino
and the strengths of the couplings of the electromagnetic, strong
and weak interactions.

We begin with point i), the demonstration of the sensitivity
of $N(p)$ to the parameters that determine low energy physics.
The sensitivity of the $N(p)$ to all parameters of cosmology and
particle physics associated with stable matter follows from the
circumstance,
mentioned in the introduction, that the existence
of main sequence stars requires a number of coincidences.  Among
these are,

1)  The existence of stable nuclei, up to at least carbon, requires
conditions on $\Delta m = m_{neutron}-m_{proton}  $, $\alpha$
and $\alpha_S$, the strong interaction coupling constant.  The
requirements are $\Delta m < 18 Mev$, that $\alpha$ not
be greater than $.1$ and that $\alpha_S $ not be weakened more
than by a factor of $2$ \cite{CarrRees,BarrowTipler}.

2)  The production of these nuclei in stars requires still stricter
limits.  An increase in $\Delta m$ by a factor of $2$ from its
present value, or an increase of $\alpha_S$ by $31 \%$, unbounds
the deuteron, while an increase in $\alpha_S$ by $13\%$ will bind
the diproton and dineutron, all of which would modify drastically
the evolution of stars\cite{CarrRees,BarrowTipler}.
Further, as first pointed out by Hoyl, that carbon is resonantly
produced, and does not resonantly burn to
oxygen, requires that the former nuclei have, and the latter not have,
a level within narrow ranges\cite{Hoyl}.
Consequently, the requirement that
carbon be produced copiously in stars is likely to put still stronger
limits on these values.

That nuclear fusion take place puts additional limits on the
parameters including\cite{CarrRees,BarrowTipler}
\f
\Delta m \approx 2 m_{electron},
\ff
\f
\alpha \approx {\Delta m \over m_\pi }
\ff
and
\f
\alpha  > {m_{electron} \over m_{proton}}
\ff
3)  Additionally, the requirement that stars that burn hydrogen
are stable and that the photon pressures contribute to, but do
not dominate, the energy balance of a star
leads to\cite{CarrRees,BarrowTipler}
\f
{m_{electron} \over m_{proton}} > \alpha^2 50^{-4/3}
\ff
\f
G_{Newton}m_{proton}^2 < \alpha^{12}
\ff

4) As Carter pointed out, the existence of convective stars requires
the more precise relationship that\cite{Carter}
\f
G_{Newton}m_{proton}^2 \approx \left
( {m_{electron} \over m_{proton} }\right )^4 \alpha^{12}
\ff
We may note that this is satisfied up to a factor of 3.

5)  The requirement that supernova exist bounds the weak coupling
constant on both sides
so that the neutrinos produced interact weakly enough to
escape the collapsing core but strongly enough so that they may
expel the envelope.  As pointed out by Carr and Rees, that
this be the case implies that\cite{CarrRees},
\f
G_{Fermi}m_{electron}^2 \approx
\left (G_{Newton}m_{electron}^2 \right )^{1 \over 4}
\left ( {m_{electron} \over m_{proton}} \right )^{1 \over 2} .
\ff

6)  If there is a grand unified gauge group, the unification
scale is restricted by the requirement that the proton lifetime
exceed the lifetime of main sequence stars
to satisfy\cite{BarrowTipler}
\f
m_{unification}  > \alpha
\left ( M_{Planck} m_{proton}     \right )^{1 \over 2}
\left (  { m_{proton}  \over m_{electron} }  \right )^{1 \over 2}
\ff

7)  We may finally note that the existence of main sequence stars
puts restrictions on all the main cosmological parameters, as has
been often discussed\cite{BarrowTipler}.

None of these relations are new, they have all been put forward
previously as evidence for the anthropic
principle, and their
derivations may be found in the cited
references\cite{CarrRees,BarrowTipler}.  What I  would
like to do here is to reinterpret each of them as evidence for
the prediction $*$.  In particular, as the small size of the
primordial density fluctuations observed by COBE \cite{COBE}, as well
as direct observational limits, seems to rule
out the presence of primordial black holes in our universe, the
dominant mode of black hole production in our
universe is by the collapse of massive stars.  As such any
change in the parameters that
effects the production or evolution of stars, or the process of
supernova, is going to effect the number of black holes.  This is
sufficient to establish the sensitivity of $N(p)$ to all of
the parameters appearing in (1)-(8).

Having established the sensitivity of $N(p)$
to the parameters that determine low energy physics, we may
go on to discuss the evidence for the conjecture $*$ in the
case of these parameters.  The evidence that changes in these
parameters in many cases descrease the number of black holes
produced
comes from the following
considerations:

i)  Black holes would not form copiously were there not galaxies.
Therefor any change in the parameters that disrupts the formation
of the galaxies will decrease the number of black holes.  While
we do not currently have a completely successful theory of
galaxy formation, it is likely that the early stages
involve the condensation of overdense regions by cooling by
bremsstrahlung processes.  That this can occur puts
conditions on $S$, the photon to baryon ratio, and the scale of
$\delta \rho / \rho $, the primordial
fluctuations\cite{CarrRees,BarrowTipler}.   For instance, it is
likely true that the formation of galaxies requires that the
decoupling
time approximately coincide with the transition from radiation to
matter dominated universe, this requires that $S \approx 10^9$, as
observed.  While it is difficult to make this more specific it is clear
that galaxies could not form in a universe with $S$ much larger than
this.

That there are electrons to bremsstrahlung requires that
\f
\Delta m >-7 Mev.
\ff
so that the universe is primordially mostly hydrogen rather than
mostly neutrons.  (The right hand side is not zero because we may
allow the possibility that if helium were stable some would
be produced in the early universe.)
We may note that there is only a factor of
$10^3$ between the cooling times of clouds of
$10^{12} M_{solar}$ and the Hubble, time, there
are then no cooling mechanisms involving only neutrons
that could play a role in galaxy formation at a time
much shorter than the present hubble
time\footnote{Rothman and Ellis\cite{EllisRothman} have studied
the proposal of cosmological natural selection and criticized the
argument that a neutron universe would be less efficient at
forming stars.  However their arguments principly apply to
the collapse of clouds to stars, whereas my point is that in
a neutron universe it is less likely that there would be many cold
dense clouds of the type that collapse to form stars. For the case
of collapse to galaxies, the electron opacity
is unlikely to slow collapse of hydrogen, while the much diminished
rate of radiation in a neutron cloud due to coupling to the neutron
dipole moments rather than electrons is likely to slow cooling
of the primordial clouds that become galaxies.
(See \cite{BarrowTipler}. eq. 6.71 for the cooling rate.)
Furthermore, the point is not whether there are ways to make
the collapse of cold clouds to stars more effecient.  These
processes in our universe {\it are} rather inefficient.  The point is
that the processes that continually form new molecular
clouds and catalyze their collapse depend on the delicate
tunings of the parameters that provide a universe copious in
carbon and supernovas. Thus,
it may be possible to change the parameters
such as to make a given cold cloud more likely to collapse to
form a star, or to make
make a given massive star more likely to be a black hole.
The problem is that such changes seem in all cases
so far studied to disrupt the
processes that are apparently necessary to have
a constant rate of massive star formation, and hence of black
hole formation.  The main claim I am making is that this constant
rate of star formation results in many more black holes than would
the episodic star formation that would result were there
not the delicate fine tunings that the present mechanisms
seem to require.}.
Thus, we may conclude that if (9) were not
satisfied,  the number of
black holes would consequently strongly decrease.

Furthermore, that galaxies are much smaller than the radius of the
universe at the time of galaxy formation, $R_{formation}$,
requires that\cite{BarrowTipler}
\f
{ \alpha^4 \over G_{Newton}m_{proton}^2 }
\left (  m_{proton} \over m_{neutron} \right )^{1 \over 2} a_{bohr}
< < R_{formation}
\ff

ii)  As black holes are
the result of the collapse of very massive,
short lived stars, it follows that a significant, and likely dominant,
mode of black hole formation in our universe is in the continual
formation of massive stars in spiral galaxies. Thus, if the
processes by which the continual process of star formation
and hence black hole formation in spiral galaxies were disrupted
by some change in the parameters, the number of black holes
produced during the lifetime of the universe would significantly
decrease, unless the same change led to a compensating increase
in the black holes formed during earlier stages of the
universe such as in the formation of elliptical
galaxies and in the halos of spiral galaxies.

It is then important to  note that recent work on spiral galaxies
has led many astrophysicists to the conclusion that star
formation in
spiral galaxies  is a
self-propagating process whose rate is likely governed by
feedback processes at several
scales\cite{spiralstructure,elmegreen-review,elmegreen-triggered}
\cite{IBMguys,parravano,elmegreen-model}.
Disruption of these
feedback processes resulting from a change of parameters
would then likely lead to a decrease in the rate of black hole
production (again, as long as there is no compensating increase
from other effects of the change.)

The evidence that self propogating star formation, with a
rate governed by feedback processes, contributes significantly
or dominantly to the star formation rate of spiral galaxies
may be summarized as follows.

1)  There is good evidence that the star formation rate in our
galaxy and other spiral galaxies is constant over the disk on
time scales of $10^{10}$ years\cite{WyseSilk}.
This is, {\it a priori},
unlikely without
self-regulation because the time scales involved in star formation
and in the significant energetic interaction between stars and
the interstellar medium range only up to $10^7$ years.  Other
evidence of this kind comes from the fact that after $10^{10}$
years the dust and gas normally constitute  a significant
fraction by mass of the disk,
between $.1$ and $.5$.  Finally,
the rate of conversion of gas and dust in the disk to stars, which
is estimated at $3-5 M_{solar} / \mbox{year}$ is
approximately  equal to the rate of return of matter to the medium
from stars, which is at
least $1-2 M_{solar} / \mbox{year}$\cite{spiralstructure}.  Given
the present uncertainties about the rates of mass loss by massive
stars and the
infall of gas into the disk from the galactic halo, it is then
plausible that the disk is in a steady state, with a lifetime of
at least a few times $10^{10}$ years\cite{WyseSilk}.  It is
important to note that this cannot be an
equilibrium state because of the enormous differences in the
temperatures and densities of the different components of the
interstellar gas; the galactic disk  is therefor a nonequilibrium
steady state system
driven by gravitational and nuclear potential energy.

2)  There are many examples in which star formation is observed
being triggered by shock waves from supernovas or the
interaction of giant molecular clouds and ionized regions heated by
massive
stars\cite{spiralstructure,elmegreen-triggered,elmegreen-review}.

3)  There is evidence that the ambient warm interstellar medium in
many galaxies is near the critical pressure and temperature for the
phase transition between warm ($100^o K$) atomic clouds and cold
($20^o K$) dense molecular
clouds\cite{parravano}.
Additional evidence that the
medium is critical is that there is good evidence that the distribution
of the cold clouds in the medium is scale invariant and fractal up to
the scales of the spiral arms \cite{scalo-fractal}.
Feedback mechanisms involving heating by massive stars have
been proposed which would keep the medium at the critical point
for this transition \cite{parravano}.

4)  There are successful models of the spiral structure
that incorporate
triggered, propagating star formation, which is regulated by
feedback effects\cite{IBMguys,elmegreen-model}.  It seems very
helpful to incorporate
such effects to achieve the  generatation of  persistant
spiral structure over a range of spiral types.  Typically, in such
models
the rate of
star formation stimulated by energetic events such as supernova
from massive stars dominates over the spontaneous rate.
These include the simple
cellular automota models of Gerola, Schulman
and Seiden\cite{IBMguys}
and more realistic models involving moving clouds
and stars by Elmgreen and Thomasson\cite{elmegreen-model}.

The cellular automata models\cite{IBMguys}
employ  directed
percolation models in $2+1$
dimensions, where the percolation probability, $p$ is tuned to be
near the critical point by feedback effects involving the interstellar
medium.  Without these feedback effects spiral structure can only
be reproduced by tuning $p$ to the percolation fixed point.
This model has further successes such as reproducing bursts
and oscillations of star formation in small galaxies, which is
observed in blue dwarf galaxies, and incorporating a natural
explanation of the lack of continual star formation in elliptical
galaxies.  It then seems likely that idealized as it is, this model
isolates the key processes of spiral structure; to the extent that
this is the case propagated star formation dominates the star
formation rate in spiral galaxies.

It is apparently the case that these percolation models have
difficulty reproducing grand design spirals.  These
symmetric patterns are reproduced by the competing density
wave theory, however that appears to have difficulty
explaining the persistance of spiral structure in isolated
spiral galaxies\cite{elmegreen-model}.
The most succesful models, such as
that of Elmegreen and Thomasson, incorporate both hydrodynamical
and feedback effects (including propogating star formation)
and are able to reproduce
persistant spiral
structure over the whole range of
spiral types\cite{elmegreen-model}.  It then seems
reasonable to conclude that the effects isolated in the
percolation models do play a role in real spiral galaxies, but
in combination with global hydrodynamical effects.

If, as the evidence seems then to
point to, the galactic disk is a nonequilibriium
system driven by gravitational and nuclear potential energy which
has evolved to a steady state in which the rate of
star formation is governed by  feedback loops, one cannot make a
simple estimate of the rate of formation of black holes as a function
of the fundamental paramters.  However another opportunity is
available to test the prediction $*$, which is that any change in the
parameters that disrupts critical processes in the star formation
process will lead to a cessation of that process and a transition to
a state in which the rate of star formation, and hence of black
hole formation, is drastically reduced.  As long as that change does
not lead to increases in some other mode of black hole formation,
one may conclude that the number of black holes formed by the
universe then significantly decreases.

There are two critical
processes involved with star formation that can be so disrupted.
These are supernovas and the transition from warm
atomic gas to the giant
molecular clouds.
We discuss them in the following two sections.

\section{Supernovas, star formation and the Fermi constant}

Type II
supernovas play a crtical role in this scenario as they are
both the events in which black
black holes are formed and the triggers for propagating star
formation\footnote{Type I supernovas are not believed to form
black holes \cite{Shapiro-T,WoosleyWeaver}.}.
As a result of the Carr-Rees observation
mentioned above\cite{CarrRees},
that type II supernovas could  not occur in a world in which the
value of $G_{Fermi}$  was either increased or decreased significantly,
we have a candidate for a substantiation of the prediction $*$.
Without supernovas there would
be no resulting shock wave to trigger star formation and also no
material returned to the interstellar medium.

This has three consequences.  First, without triggered star
formation the scenario discussed in the previous section implies
that the rate of star formation, and hence of black hole
formation significantly decreases.  Furthermore,
whatever star formation rate persists in this case, there is less
material available for the formation of new stars, as there is no
return of matter to the interstellar medium from supernova.
Third, those massive stars that are formed are more likely to
form black holes, as, without supernova, the envelope would remain
bound to the core, resulting in the colapse of the whole massive
star to a black hole.

However, while the number of massive stars that, once formed, became
black holes would certainly increase in this case, the issue is
how many massive stars a universe without supernova would
 form to begin with.
It is certainly plausible that the answer is
a great many fewer.  The reason
is that the formation of very massive stars requires very energetic
events which can force the clouds of gas and dust to sufficient
densities that gravitational collapse can overcome the thermal
and magnetic support of the clouds.
The very low effeciency of the
star formation process attests to the apparent fact that the rate for
this to occur spontaneously is low.

Furthermore, this is in fact most
likely to be case for massive stars, because it is correspondingly
less likely, in the absense of violent events such as shock waves,
for the clouds to collapse sufficiently fast for masses many times
the Chandrasekar mass to accrete before the process is reversed by
winds driven by processes in the protostar.  These processes are
quite effecient at halting most cloud collapses shortly after
the protostar ignites, as is evidenced both by the fact that
most stars that form are small and by the low effeciency of
the conversion of the mass of giant molecular clouds into
stars.  The evidence for there
being a bimodal initial mass function\cite{Larson,Scalo},
as well as for massive stars
forming in distinct regions\cite{stars} attests to this.

Thus, it is reasonble to conclude that it is likely that the rate of
spontaneous formation of massive stars is very small, so that in the
absence of supernovas very few of these stars would be formed.  This
effect may then overwhelm the fact that in such a world more of the
massive stars that did form would become black holes.

It may seem novel that important astrophysical processes
depend on fine tunings of the parameters of particle physics.
It is interesting that it is not hard to find a rather general
argument that this may be the case.   To give this we will
assume that the star formation rate $R(t)$, where we have
indicated its possible dependence on time, is the sum of a
spontaneous process and a process driven by supernovas so that
\f
R(t)=A +B S(t)
\ff
 where $A$ gives the spontaneous rate, $S(t)$ is the supernova rate
and $B$ is the number of new stars whose formation is induced by
each supernova.  We may assume that the supernova rate is given
by
\f
S(t) = R(t-\tau_{sn}) \int_{m_{sn}}^\infty dm D(m)
\ff
where $\tau_{sn}$ is the average time from formation to supernova
of a massive star and $D(m)$ is the initial mass function, which
is defined so that $D(m)dm$ is
equal to the proportion of stars that form with masses
between $m$ and $m+dm$.  I have here normalized it so that
$\int_0^\infty dm D(m) = 1$.  We may assume
that $D(m)$ is zero below some lower mass cutoff which is less
than $m_{sn}$, which is the minimal mass that results in a
supernova.  Above this we assume it takes the simple form
$D(m) = (\beta -1) /m_{0} (m/m_{0})^{-\beta}$, where the parameter
$\beta$ is known to be greater than one.
One then easily finds that
\f
R(t)=A +  B  R(t-\tau_{sn} )
\left ( { m_0 \over  m_{sn}} \right )^{\beta - 1}
\ff
Thus, if the star formation rate is constant, as is observed, we have,
\f
R=R(t) = {  A \over {1 - B
\left ( { m_0 \over  m_{sn}} \right )^{\beta - 1} }}
\ff
if $A \neq 0$ or
\f
B= \left ( { m_{sn} \over  m_{0} }\right )^{\beta - 1}
\ff
if there is no spontaneous star formation.   Now, both observation
and the success of the stochastic models of spiral structure
suggest that there is a small spontaneous star formation rate, but
that the dominant process is induced star formation triggered by
supernova bursts.  If this is the case, and if, as we assumed, the
star formation rate is constant, this requires that the constant
$B$ be tuned so that the equality (15) approximately hold.

As $B$ is the number of star formation events induced by a single
supernova, it is sensitive to the energy created by each supernova
and hence to the weak coupling constant.  This argument shows that,
given the assumptions, the value of $G_{Fermi}$ falls into a narrow
range that allows a constant rate of induced star formation to
dominate the star formation process of the galaxy.  To put this
another way, the spiral appearance of the galaxies may be regarded
as the result of the weak coupling constant being tuned so that (15)
approximately holds.

To conclude the argument, it is necessary to check that increases
or decreases in $G_{Fermi}$ large enough to suppress type II
supernovas do not lead to other mechanisms for black hole
formation.  One important side effect that must be considered is
that the fact that some, but not all, of the baryons are bound into
helium depends also on the
coincidence (6)\cite{CarrRees,BarrowTipler}.  Thus, an increase in
$G_{Fermi}$ leads to a world that is all hydrogen primordially,
while a decrease
will lead to a world that is primordially all helium.  It is difficult
to imagine that an all hydrogen world would have drastically
different rates of star formation and black hole formation than
our universe, but the case of a helium universe is more difficult.
One effect would be that all stars would now have lifetimes of
$10^{6-7}$ years.  The result could be an increase in the rate of
type I supernovas, as there would be a much larger number of
white dwarfs formed within the hubble time.  However, it is
generally believed that type I supernovas do not lead to black
holes.  A more difficult question, which is so far unresolved, is
whether the intitial mass function might increase on the high
mass side in a helium world.

This ends the argument that small changes in $G_{Fermi}$
may plausibly lead to
decreases of the rate of black hole production in spiral galaxies,
in agreement with $*$.

\section{Star formation and carbon}

The second critical process in spiral galaxies is the cooling of
the dense molecular clouds, leading to star formation.  A scenario
for this process that seems consistent with observations to
date is the following\cite{stars}.
Dense molecular clouds form spontaneously
in the interstellar medium as a result of cooling processes involving
dust.  Star formation then occurs in these clouds by further
condensation of small regions of the clouds. The process by which
stars are formed from the dense molecular clouds is not very
efficient, possibly because the clouds are supported by magnetic
fields, so that the overall efficiency of conversion of clouds into
stars is about one
percent\cite{stars,WyseSilk}.  Because of this, induced
processes, in which the collapse of parts of the cloud are catalyzed
by shock waves from supernova, make an important contribution
to the star formation rate, in addition to whatever spontaneous
rate of star formation may exist.

Thus, in addition to supernovas, the processes by which the dense
molecular clouds cool and condense are critical for there to be a
constant rate of star formation, and hence black hole
formation, in spiral galaxies.  We may note that both the
dominant cooling
mechanisms of the clouds and the shielding of the interiors of the
clouds to heating from ultraviolet radiation from young stars
require the presence of carbon, in the form of dust and in
the form of CO, whose transitions provide the dominant
cooling.  (Furthermore, it is possible that the CO and other
molecules are formed on the surface of the dust.)   Therefor,
we may conclude that any change in the parameters of particle
physics that results in carbon nuclei being either unstable or
not copiously produced in stars will lead to a decrease in the
rate of formation of black holes, because there would not be
possible a constant rate of star formation over the life of the
galaxy.

If we recall the arguments of section 3 we will see that the
requirement that the carbon nuclei be both stable an copiously
produced puts strong constraints on many of the parameters,
from equations (1-8).  We may then conclude that small changes
in all of these parameters that lead to violations of these relations
will result in a decrease in the number of black holes produced
by spiral galaxies, and, hence, by our universe.

\section{Some further tests of the conjecture}

Given the a priori implausibility of the conjecture $*$, it is surprising
that it is not possible to discover many changes in the paramters of
physics and cosmology that lead to strong increases in the number
of black holes produced by the universe.   Indeed, as several people
have pointed out, there are several candidates for such
changes that come immediately to mind.  I would like to devote this
next to last section of this paper to discussing them and explaining
why they do not immediately lead to a refutation of the
 conjecture $*$.     At the same time, in at least two of
 the cases, there is a possibility that more work will
reveal that the conjecture is refuted.  These are then
clearly important directions for further work.

\subsection{Increasing the gravitational constant}

One change that might seem to lead to the formation of more black
holes is to increase the strength of the gravitational force.  Surely by
hastening gravitational collapse more black holes will be created.

However, when looked at more closely it is not at all obvious that to
increase $G_{Newton}$ will lead to an increase in the number of
black holes.  The main reason is that the mass of a typical star
scales as the same power of $G_{Newton}m_{proton}^2$ as does the
Chandrasekar mass,
$M_{Chandra} \approx m_{proton} (G_{Newton}m_{proton}^2)^{3 \over 2} $,
to which the upper limit for the mass of a
stable neutron star is proportional\cite{CarrRees,BarrowTipler}.
The reason is believed to be
that the process of collapse of a dense core of a giant molecular
cloud to a star is halted by energy released by the ignition of nuclear
fusion\cite{stars}, which happens at a
mass proportional also to
$M_{Chandra}$.  Thus, the main effect of increasing $G$ will be
to make {\it all} stars proportionately more massive, but it would
not directly change the proportion of stars that become black holes.

Furthermore, if the mass available in a galaxy or in the whole
universe to be turned into stars is fixed, then an increase in the mass
of each star would lead to a {\it decrease}
in the number of total stars
and, if their proportion is unchanged, to a {\it decrease}
in the number of
black holes.  We may note that as $M_{Chandra}$ increases like the
$3/2$ power of $G_{Newton}m_{proton}^2$, this effect could be very
significant.

Secondly, increasing $G$ significantly will make all stars unstable
because (5) is then violated, while even modest increases in $G$
will change stellar evolution significantly because (6) is violated.

A third effect of increasing $G_{Newton}m_{proton}^2$ would be to
strongly decrease the lifetime of each kind of star, which is
proportional to $(G_{Newton}m_{proton}^2)^{-2}$.  However, the
collapse times for clouds of dust and gas, on which depend the time
scales for the processes of star formation are proportional
to $(G_{Newton}^{5 \over 2} $   \cite{EllisRothman}.
This means that an
increase in $G_{Newton}m_{proton}^2$ will quickly lead to a
situation in which the life time of a massive star, from birth to
supernova will be the same as the time scale of star formation.
This will disrupt the processes of star formation because no
giant molecular cloud would be able to form more than a few stars
before it would be disrupted by a supernova, drastically reducing the
efficiency for the formation of gas to stars, and hence decreasing the
star formation rate.

While these processes are complex enough that it is difficult to
draw definitive conclusions, it seems that there is no reason to
expect that an increase in $G_{Newton}m_{proton}^2$ will lead to
a decrease in the rate of formation of black holes and
several pieces of evidence that it would
have the opposite effect.

\subsection{Increasing the number of baryons}

A commonsense way to increase the number of black holes would
be to increase the amount of matter available to form stars and
black holes.   However,
as we do not know if our universe is finite or infinite, we do
not know if we can speak of a total number of baryons
in the universe.  But it certainly does make sense to speak of
increasing the proportion of matter that is in baryons.  If we
otherwise keep the history of the universe fixed, this has the
effect of decreasing the photon to baryon ratio $S$.

Decreasing $S$ greatly affects the history of the early universe,
necessitating changes in the scenarios for nucleosynthesis and
structure formation.   Cosmological
scenarios in which $S$ is intially much lower, called cold or tepid
big bang models\cite{coldbigbang}, have been
studied, and it is possible to
arrive at the same proportion of helium as in our present
universe\cite{nucleosynthesis}.    The main issue with
such a scenario is whether there
are viable scenarios for structure formation, leading to galaxies and
hence to black holes.

At the same time, it may not be that $S$ is a free parameter.  If
it arises instead from $CP$ violating effects in the early universe
then $S$ is inversely proportional to the $CP$
violating\cite{BarrowTipler}.  To decrease in this case
$S$ then requires that $CP$ violating effects are increased.
Such a change is unlikely to affect the properties of ordinary matter.
Thus,
if the problem of structure formation can be solved, this is a
candidate for violation of $*$  that deserves further exploration.

\subsection{Lowering the upper mass limit for neutron stars}

A change that would certainly lead to an increase in the number of
black holes would be a decrease in the upper mass limit for
neutron stars.  This would lower the mass needed to form a black
hole, which would result in the formation of more black holes.

The difficulty is that the upper mass limit for neutron
stars depends only on the Chandrasekhar mass and the equation
of state for
nuclear matter\cite{Shapiro-T}.   It is certainly possible
to lower
the upper mass limit by changing from a stiffer to a softer
equation of state.  However, the  physics that dominates the
 determination of the equation of state for nuclear matter is
$QCD$, which has no free parameters apart from the dimensional
$QCD$ scale and the quark masses.  A change in  these
 parameters might achieve a softer equation of state, but there
will be other effects on the rates of key processes involved in
stellar physics.  These are likely to strongly effect in other ways the
number of stars and black holes produced.    In particular, as the
present formation of black holes depends on the several coincidences
we have already discussed, it is not clear if the equation of state
could be softened without disrupting the processes that lead to
constant star formation rates in galaxies.

However, it cannot be ruled out that there is a change
in some of the parameters of nuclear physics that will
soften the equation of state while leaving unaffected the
binding of deuterium and the ability of stars to produce
carbon copiously.  One interesting such possibility is that
this might be accomplished by changing the strange quark
mass, as it has been conjectured that neutron stars have a
significant component of strange matter.  Thus, this is a
possibility that deserves further exploration.

\subsection{Changing the slope of the initial mass function}

Another obvious way to increase the numbers of black
holes produced would be to increase the proportion of the
material of the galactic disk that is made into massive stars,
in relation to the proportion that is made into small stars.
Such a change
would have a two fold effect on the final number of black
holes produced, first because more massive stars are made
at one time and second because most of the matter that goes
into massive stars that supernova is recycled back into the
interstellar medium, while a smaller proportion of the matter
that goes into smaller stars is recycled.  (Although it should be
mentioned that the proportion of matter recycled due to steller
winds from stars is believed now to be the significant
contribution to recycling, dominating over the mass remnants
of supernovas.  Further, the present rate of recycling of matter
is not small, it is estimated to be about $40 \% $ in the
solar neighborhood\cite{WyseSilk}.)

The proportion of matter going into massive stars is determined
by the shape of the initial mass function, which is believed to
follow a power law for large
masses\cite{Salpeter,Scalo}.   Unfortunately, for large
masses that are relevant for this question, that power is only
poorly measured.  Doubly unfortunately, we do not understand
the physics that determines what the slope of the initial mass
function is.    For example, it is not even agreed upon whether
there is a single process that produces stars of all masses, or
two different processes, one of which produces low mass stars,
while the other is predominantly responsible for the production
of massive stars\cite{Larson,Scalo,stars}.

This is then also a subject that deserves further work.
There is only one point
which might be mentioned, which is that if it is the case, as
present evidence seems to suggest, that the rate at which material
is formed into stars is matched, in spiral galaxies,
by the rate of the return of material from stars to the interstellar
medium, then this matching must be sensitively dependent on the
slope of the initial mass function.  This leads to two possible
conclusions, first that changes in the slope of the initial mass
function will disrupt this balance, making the continual star
formation-and hence black hole formation-of spiral galaxies
impossible.  The result will either be no star formation as in the
elliptical galaxies, or a temporary runaway star formation as in
the star burst galaxies.

The second conclusion is that it may be that the relative proportion
of low mass and high mass stars is itself determined by some process
of self-regulation that effectuates the balence between the rate
of mass flow in each direction between stars and the interstellar
medium.  This is not impossible, especially if a separate process is
responsible for the formation of high mass stars.

For example if
the process of self-propogating star formation, through
supernova caused shock waves is primarily responsible for
the formation of massive stars, as has been
proposed\cite{stars}, then there
is a natural feedback process that adjusts the  rate
of this process to the amount of
material available in giant molecular
clouds\cite{IBMguys,spiralstructure,elmegreen-review}
\cite{elmegreen-triggered}.  Too much star
formation depletes the interstellar medium, making subsequent
supernova shocks less effecient in catalyzing the formation
of new stars.  But too little star formation results in the
collection of more clouds, making subsequent supernova
shocks more efficient as catalysts of new star formation.
Such a feedback
mechanism is, indeed, essential to the models of spiral structure
of Gerola, Seiden and Schulman\cite{IBMguys}.

The point, beyond the simple beauty of such possible
mechanisms, is that
if this is the case there is no parameter that can be varied to
increase the proportion of matter that goes into massive stars
and hence black holes.  An imagined galaxy that would produce
many more black holes in each generation of star formation
could not support a constant rate of
star formation, hence the overall black hole formation rate would
decrease.

\subsection{Early production of black holes}

Notwithstanding what has just been said, it has sometimes
been conjectured that the relative proportion of massive and
light stars does change in time, with a higher proportion of
massive stars produced at earlier times\cite{Larson,Scalo}.
A possible reason
for this might be that a certain enrichment of the interstellar
medium with carbon and other elements is necessary for the
mechanisms of the formation of light stars that we see now,
which is dominated by cooling of giant molecular clouds involving
such metals.  We may note that it is only such slow, regulated,
mechanisms of star formation that can produce
stars predominantly around a solar mass, as the collapse has
to be easily reversed soon after nuclear ignition has taken
place in the center of the protostar.

At earlier times, before the medium was enriched,
it may be that the only available mechaisms for star formation
were more violent, with shocks from supernovas playing a more
important role.  It has then been conjectured that in the early
history of a galaxy many more massive stars were formed,
in what might have been runaway chain reactions of
massive star formation and supernova
explosions\cite{JaneSalpeter}.  The result,
beyond the enrichment of the medium to the point that
formation of light stars through cooling became possible, would
be that a significant portion of the halos of galaxies may be in
relic neutron stars and black holes from this period.

If this is the case then such early processes might make a
significant contribution to the total black hole production
of a galaxy.  Again, this is a question that deserves further
exploration.

It has also been suggested that shortly after decoupling there was
a burst of massive star formation, which resulted in the formation
of a large number of black holes, which would presently constitute
a major proportion of the dark matter and
inside of which a large fraction of
the baryons would be trapped\cite{GnedinOstriker}.  This possibility
is consistent as well with the recent
observations\cite{Omega,GnedinOstriker}
that point to a
value of
$\Omega = .1-.2$.  Such early processes would contribute
significantly to the black hole production of a universe and also
deserve further exploration in relation to the conjecture $*$.

\subsection{The issue of $\Omega$}

Finally, there is the question of the density of matter, and the
value of $\Omega$.  As is well known, theories that $\Omega$
is determined by elementary particle physics, such as inflationary
models, predict uniformly that $\Omega$ should be equal to one.
The general argument for $\Omega =1$ is simply one of
scales; if it has any other value then there is a dimensional
parameter, $\tau_{universe}$,
which is the lifetime of the universe before it either
recollapses or becomes very dilute.  The fact that this has
not yet happenned means that this parameter is at least
as great as several times the present age of the universe.  The
great mystery is then why the laws of elementary particle
physics that governed the early universe should produce such
a parameter, which is enormously greater than the natural time
scales of elementary particle physics.  The difficulty of answering
this question results in the natural expectation that there is no
such parameter, which is only possible if $\Omega = 1$.

It should then be mentioned that the scenario of cosmological
natural selection discussed here does provide a natural
explanation for $\tau_{universe}$ being several times the present
age of the universe.  The reason is simply that if such a parameter
were fixed by the conjectured process of cosmological natural
selection, we would expect it to be not significantly
longer than the time scale
over which galaxies produced significant numbers of black holes.
While the rate of star formation is approximately constant in
spiral galaxies, there is evidence that the rate is decreasing
on scales of $10^{9-10}$ years, coming from both the observations
of many blue galaxies at high redshifts and models of chemical
evolution of the galaxy\cite{WyseSilk}.
If this is the case then there may be a
time on the order of perhaps ten times the current age of the
universe at which the rate of formation of black holes has
strongly decreased.  If this is the case then, on the scenario of
cosmological natural selection, we would expect the overall
lifetime of the universe to be not significantly greater than this
time.

While this is very rough, given present knowledge, we may note that
this would result in an $\Omega$ presently of not $1$, but
more likely around $.1$.  It is interesting to note that, while there
are not yet conclusive results, the value of $.1-.2$ is what is
claimed by observational astronomers\cite{Omega,GnedinOstriker}
as
the most likely value
for $\Omega$.

Further, we may note that if the parameters of cosmology and
particle physics have been tuned by a random and stochastic
process such as cosmological natural selection, it is more likely
that the effect that extermizes the production of black holes is
produced by tuning several parameters that effect the result
equally roughly, rather then tuning one or more
of them extremely
finely.  As the cosmological constant, the neutrino
mass, as well as the initial mass density all contribute to
$\Omega$, if this scenario is true we should then expect that
the value of $\Omega$ that maximizes black hole
production is achieved through a simultaneous tuning of all
these parameters.  This would mean that we would expect to
see a small cosmological constant, a small neutrino mass, making
some contribution to the dark matter, {\it and} at the same time
$\Omega$ on the order of $.1-.2$.

To avoid confusion I should mention that the scenario of cosmological
natural selection is compatible with inflation.  Indeed as
was discussed in \cite{evolve} it may also explain how it is
that the self-coupling of the inflaton field, $\lambda$,
is tuned to the unnaturally
small values requried for inflation.  But, especially given that the
initial density perturbations are also proportional to the same
coupling, the mechanism should tune the value of $\lambda$
to values small enough to cause sufficient inflation for a universe
like ours to be created, but there is no reason for the tuning to
be better than this.  This again leads to the conclusion that even
if there is inflation it did not last long enough to tune $\Omega$
presently any closer to one than would be required for the universe
to live as long as galaxies produce black holes.

As this differs substantially from the prediction of
conventional inflationary
models, we may regard the measurement of $\Omega$ as a test
that distinguishes the theory described here from other possible
explanations of how the cosmological parameters came to be so finely
tuned.

\section{Conclusion}

Putting these arguments together,
we see that there is good evidence that
the following changes in the parameters will lead to a decrease
in the number of black holes produced in spiral galaxies in our
universe:  i)  A reversal of the sign of $\Delta m$.  ii)  An increase
or decrease in $G_{Fermi}$ large enough to effect the energy
and matter ejected by supernovas.  iii)  An increase in
$\Delta m=m_{neutron}-m_{proton}$,
the electron mass, the neutrino mass, $\alpha$ or a decrease in
$\alpha_{strong}$
large enough to destabalize carbon (or any simultaneous change
that has the same effect).   In addition to this, the same effect will
follow from any (unfortunately unknown) changes in the parameters
that result in the coincidence of nuclear levels that are, as
noted by Hoyle, necessary for carbon to be copiously produced in
stars\cite{Hoyl}.

In addition to this, it is likely that there are further relations that
may be implied by $*$ that may emerge from a more detailed
understanding of stellar physics and cosmology.
These include bounds that follow
from the Carter relation (6) and changes in $\alpha$ and
$m_{electron}/m_{proton}$ that effect the rates of critical processes
in star formation and evolution
as well as relations that could bound $S$ and
$\delta \rho /\rho $ that may come from an understanding of
galaxy formation.   There are, however, some
open possibilities which should be further explored, among these are
the effect of changing the strange quark mass
on the equation of state for nuclear matter and hence on the
upper mass limit for neutron stars.

Finally, it should be mentioned that such a cosmological
scenario can predict why a natural time scale for the evolution of
the universe should be the time over which spiral galaxies
continue to copiously produce new stars.  This is consistent
with present observational suggestions that $\Omega = .1-.2$.
It is then very interesting that a conjecture that ties together
the large scale parameters of cosmology with the question of the
determination of the parameters of the standard model of
elementary particle physics can predict values for $\Omega$
different from $1$.

In conclusion, the conjecture
$*$ leads to, and is verified by,
 a surprisingly large number of relations among the
observed values of the
fundamental parameters of particle physics and cosmology.  If there
were really no relation between the fundamental parameters of
elementary particle physics and the rate of production of black holes,
it seems that it ought to be easy to discover ways to change the
constants to strongly increase the number of black holes.  The fact
that it seems difficult to do this suggests, at the least, that
in spite of the unusual nature of the cosmological scenario that
implies it, this conjecture may be considered to be
deserving of further development and testing.

\section*{ACKNOWLEDGEMENTS}

I am greatful to a Jane Charlton, Louis Crane, Eric Feigelson,
Ted Jacobson,
Carlo Rovelli, Peter Saulson and Larry Schulman for many discussions
about these issues.
A conversation with
Frank Shu was also extremely helpful.  The chance
to present these ideas to a seminar at the Institute for Theoretical
Physics in Santa Barbara was very helpful, and led to several
suggestions which have been explored here.
This work was supported by the National
Science Foundation under grants PHY90-16733 and
INT88-15209 and
by research funds provided by Syracuse University
and the Pennsulvania State University.

\end{document}